\documentclass[referee]{cjaa}           

\usepackage{graphicx}                   
\input{epsf.sty}                        
\input{psfig.sty}                       

\newcommand{\RA}[3]{{#1}^{{\rm h}}{#2}^{{\rm m}}{#3}^{{\rm s}}}
\newcommand{\Dec}[3]{{#1}^{\circ}{#2}'{#3}''}
\newcommand{\gsim}{\mbox{\hspace{.2em}\raisebox{.5ex}{$>$}\hspace{-.8em}\raisebox{-.5ex}{$\sim$}\hspace{.2em}}}

\begin{document}

     \title{Narrow Band {\sl Chandra} X-ray Analysis of
            Supernova Remnant 3C 391}


   \author{Yang Su
      \mailto{}
   \and Yang Chen
      }
   \offprints{Yang Su}                   

   \institute{Department of Astronomy, Nanjing University, Nanjing 210093,
              China\\
             \email{ysu@nju.edu.cn}
          }

  \date{Received~~2004 month day; accepted~~2004~~month day}

   \abstract{
We present the narrow-band and the equivalent width (EW) images of
the thermal composite supernova remnant (SNR) 3C 391 for the X-ray
emission lines of elements Mg, Si \& S using the {\sl Chandra} ACIS
Observational data. These EW images reveal the spatial
distribution of the emission of the metal species Mg, Si \& S in
the remnant. They have clumpy structure similar to that seen from
the broadband diffuse emission, suggesting that they are largely
of interstellar origin.
We find an interesting finger-like feature protruding outside the
southwestern radio border of the remnant,
which is somewhat similar to the jet-like Si structure found in
the famous SNR Cas~A.
This feature may possibly be the debris of the jet of ejecta
which implies an asymmetrical supernova explosion of a massive progenitor star.
   }
   \authorrunning{Y. Su \& Y. Chen }            
   \titlerunning{Narrow Band {\sl Chandra} X-ray Analysis of
            Supernova Remnant 3C 391 }  

   \maketitle
%
%

\section{Introduction}           
\label{sect:intro}
  3C 391 (G31.9+0.0) is an irregular
mixed-morphology (Rho \& Peter 1996, Chen \& Slane 2001) supernova
remnant (SNR) which generates bright thermal X-ray emission
interior and has a faint X-ray rim. The remarkable radio shell
which extends from the northwest (NW) to the southeast (SE) shows
that the SNR has broken out of a dense region into an adjacent
region on lower density (Reynolds \& Moffett 1993). 3C 391 is
similar to W28, W44, IC 443 and other SNRs (Reach \& Rho 1998) in
that they interact with molecular clouds, characterized by the
hydroxyl radical emission (Green et al.\ 1997; Yusef-Zadeh et al.\
2003).

We observe SNR 3C~391 with the Advanced CCD Imaging Spectrometer
(ACIS) on board the {\sl Chandra} observatory and unveil a highly
clumpy structure in
broadband (0.3-1.5, 1.5-3, 3-7, and 0.3-7 keV) X-rays (Chen et
al.\ 2004, hereafter CSSW04). 
The spectrum analysis favors a solar
abundance for the X-ray emitting gas. In the broadband images, a
faint emission of gas seems to spill out of the south-western (SW)
radio border which corresponds to the blast wave there. The nature
of such a spill out leakage of gas is unclear.

Here we report the results from the narrow-band and the equivalent
width (EW) images of the X-ray emission lines for metal species
Mg, Si and S using the observation of the ACIS detector on
board the {\sl Chandra} X-ray Observatory.
The EW maps for the element
species Mg, Si, and S allow us to distinguish the more detailed
structure by the less continuum contamination, and 
might also be useful to understand the line emission distributions
and variations throughout the remnant generally.
The narrow-band
images and the EW images are similar to the whole broadband (0.3-7
keV) X-ray emission in the overall elongated morphology. 
The narrow-band and
EW images reveal, apart from the clumpy structure similar to that
seen in the broadband image, a finger-like protrusion in Si \& S
lines outside the SW radio border. The physical significance of
this protrusion then is discussed.



\section{Data}
\label{sect:data}


Here we revisit the {\sl Chandra} ACIS observational data of SNR
3C 391 (ObsID 2786) with an exposure of 61.5 ks, which was used by CSSW04 for
spatially resolved spectroscopic analysis. The observation was carried out
with the ACIS-S3 CCD chip using very faint mode.
The level 1 raw event data were reduced to generate a level 2
event file using standard threads in the {\sl Chandra}
Interactive Analysis of Observations (CIAO) software package
version 3.1. During the course of the reprocessing,
we filtered
bad grades, applied good time intervals (GTI) correction, and
removed significant background flares to reduce the
contamination. Also, the overall light curve was examined
for possible contamination from time-variable background.
The reduced data, with the effective exposure of 60.6 ks,
are used for subsequent narrow-band analysis.


\section{Narrow-band and Equivalent-Width Images}
\label{sect:narrow}

%
\begin{figure}
   \vspace{2mm}
   \begin{center}
   \hspace{3mm}\psfig{figure=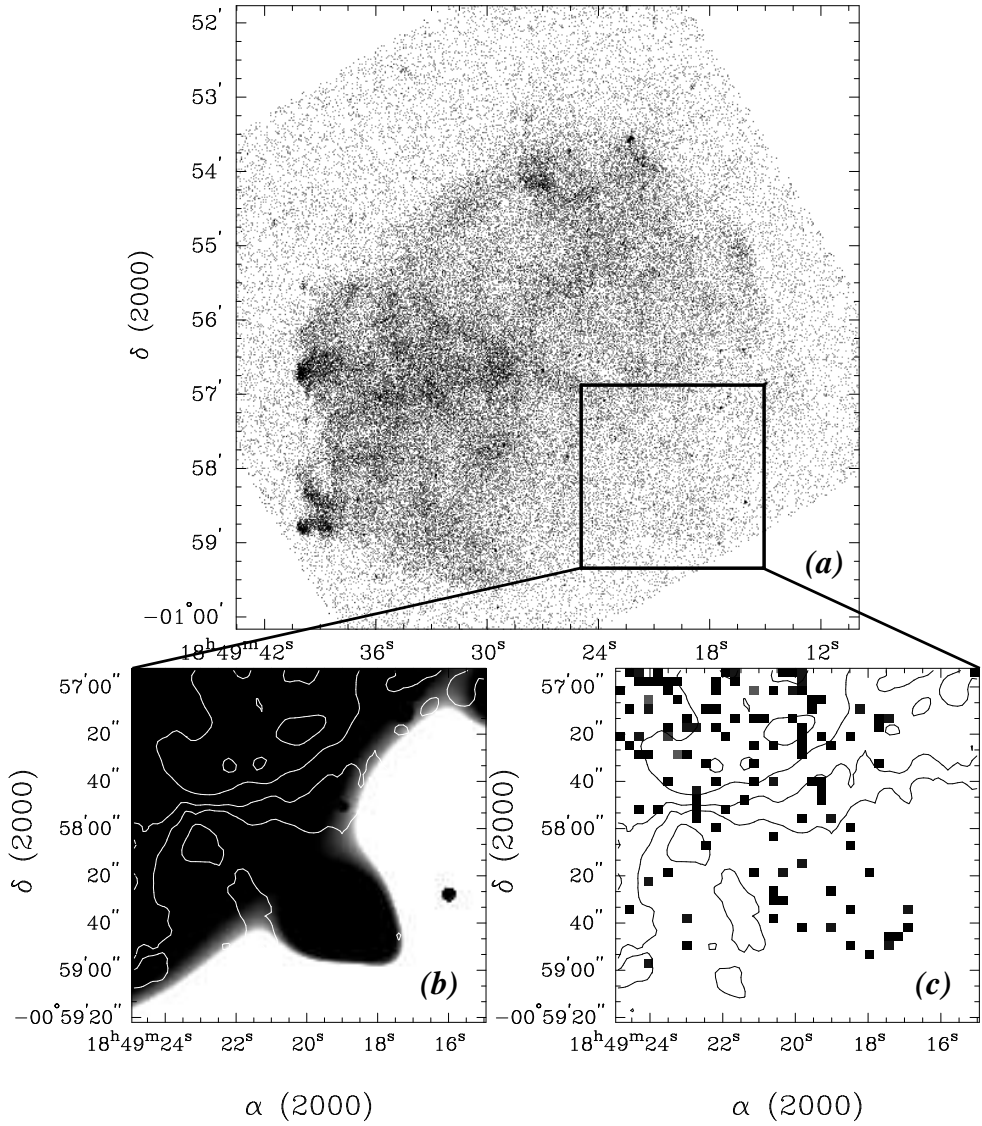,width=130mm,height=148.889mm,angle=0.0}
   \caption{
  (a). Broadband (0.3-7 keV) ACIS S3 raw image of 3C 391. The SW box
($2.5' \times 2.5'$) indicates the region of the spill-out emission (CSSW04).
  (b). Adaptively smoothed broadband (0.3-7 keV) image (with S/N ratio of 3)
of the SW region. The overlaid 1.5 GHz radio solid
  contours are at 1.5, 3.19, 8.25, and $16.69 \times 10^{-3}$
  Jy/beam (from Moffett \& Reynolds 1994).
  (c). Unsmoothed line-to-continuum ratio (i.e., EW) image for Si in the same
region as in (b), overlaid with
the 1.5 GHz radio contours with the same levels as in (b).
All the X-ray maps used here are exposure-corrected and plotted with
a square-root scaling.
 }
      \label{Fig:3C 391 raw image}
   \end{center}
\end{figure}

We present the exposure-corrected broadband (0.3-7 keV) raw image
of 3C 391 in Fig.~1a, in which many clumpy features as discussed
by CSSW04 can be discerned. The overexposed broadband image of the
SW region shown in Fig.\ 1b highlights the spill-out X-ray
emitting gas mentioned in CSSW04.

In the {\sl Chandra} ACIS spectrum (1-3 keV) of 3C 391  (Fig.\ 2),
one can see that substantial X-ray emission comes from metal
 lines, such as Mg He$\alpha$ (1.35 keV), Si He$\alpha$ (1.85 keV),
 and S He$\alpha$ (2.46 keV). With respect to the broadband map,
the narrow-band and the EW line-emission maps will show us the
morphological and positional details of the emissions from various
metal species.

The narrow-band Mg He$\alpha$(1.2-1.5 keV), Si He$\alpha$(1.7-2.0
keV), and S He$\alpha$(2.3-2.6 keV) images (without subtraction of
the continuum contribution) is given in Fig.\ 3. These three
narrow-band images have been adaptively smoothed (using CIAO tool
{\em csmooth} with signal-to-noise ratio of 3) and exposure-corrected.
The clumpy structure in narrow-band is similar to that seen
in the broadband images (CSSW04). Comparing these narrow-band
images with the overlaid 1.5 GHz radio contours (from Moffett \&
Reynolds 1994), we find that the Si and S images (Fig.\ 3b and 3c)
show a finger-like protrusion, positional consistent with the
broadband ``spill-out'' emission (Fig.\ 1b) in the southwest.

Next, we want to see the spatial distribution of the emission
purely of metal species Mg, Si, and S.
Because X-ray line and thermal continuum emissivities are both proportional to
the emission measure, the line emission in a region of low surface
brightness may still be strong relative to the continuum.
The line emission map with the correction for underlying continuum
may help to reveal the spatial distribution of metal species.
Therefore we produced the EW maps, essentially following the methods and steps described
by Hwang et al.\ (2000) and Park et al.\ (2002, 2003).

We selected photons corresponding to each particular spectral line
by identifying appropriate energy bands, as shown in table
1. In mapping the line emission, we subtracted the underlying
continuum from each line image and then get a ratio image relative
to the continuum. Since Mg He$\alpha$ line is blended with Fe~L
complex, we combine them together to increase statistic photons.
First, we extracted images in two narrower energy bands on
the both sides of each line profile (Fig.~2) and minimize the
contamination from the line emission. The underlying continuum was
calculated by linearly interpolating from the two sides. The
estimated continuum flux was integrated over the selected line
width and subtracted from the line emission. Then, the
continuum-subtracted line in intensity was divided by the
estimated continuum on a pixel-by-pixel basis to generate EW
images for each element. In order to avoid noise in this process
and decrease the poor photon statistic uncertainty, we have set EW
values to zero where the integrated continuum flux is greater than
the line flux and the intensity of the result is too low or too
high. The EW images for Mg and Si were constructed with 4$''$
pixels and smoothed by a Gaussian with $\sigma=$12$''$, and that for
S was constructed with 8$''$ pixels and smoothed by a Gaussian
with $\sigma=$16$''$. The gauss smoothed EW images are shown in
Fig.~4. We note  that the Si EW image is relatively statistically
stronger than the other two because of the high X-ray intensity,
and the S map is slightly weak for the low photon statistics.

The EW images for Mg, Si \& S lines display again the SE-NW
elongated morphology which is also seen in the broadband and
narrow-band images. The metal emission appears clumpy, too, with
the gas structure seen in the broad- and narrow-band images. Many
bright knots or clumps appearing in the broadband images (CSSW04),
such as those located at around
    $(\RA{18}{49}{39}.4,\Dec{-00}{58}{38})$ (region \#1),
    $(\RA{18}{49}{39}.6,\Dec{-00}{56}{43})$ (region \#2),
    $(\RA{18}{49}{27}.3,\Dec{-00}{54}{06})$ (region \#3),
    $(\RA{18}{49}{29}.3,\Dec{-00}{56}{42})$ (region \#6),
    $(\RA{18}{49}{32}.9,\Dec{-00}{56}{58})$ (region \#7),
    $(\RA{18}{49}{35}.5,\Dec{-00}{57}{04})$ (region \#8), and
    $(\RA{18}{49}{29}.8,\Dec{-00}{57}{45})$ (region \#9),
have their counterparts in the EW images. All these knots and clumps
have been found to be of solar abundance (CSSW04), therefore the clumpy
line emission revealed by the EW maps would arise from the clumpy
interstellar medium (ISM).

It is interesting that the SW finger-like feature appears clearly
in the EW maps of Si and S lines (Fig.~4b, 4c). 
A very faint hint of the Mg emission at the site of the protrusion 
is marginally discerned (Fig.~4a).
Furthermore, the
shape of the ``finger'' in the image for Si (Fig.~4b) bears a
close resemblance to that in the smoothed broadband image (Fig.~
1b). For comparison, we display the unsmoothed EW image for Si
including the SW finger-like feature in Fig.~1c. In Fig.~1c,
this protruding structure is measured to extend about 1.3$'$
(approximately 3 pc at a distance 8 kpc) outside the radio
shell. We note that the EW maps depend not only on the element
abundances but also on the temperature and on the ionization age,
so that the given EW images should only be taken as an indication
of some feature which represent the position and brightness of
line emission but not the actual maps of the mass distribution.

\begin{figure}
   \vspace{2mm}
   \begin{center}
   \hspace{3mm}\psfig{figure=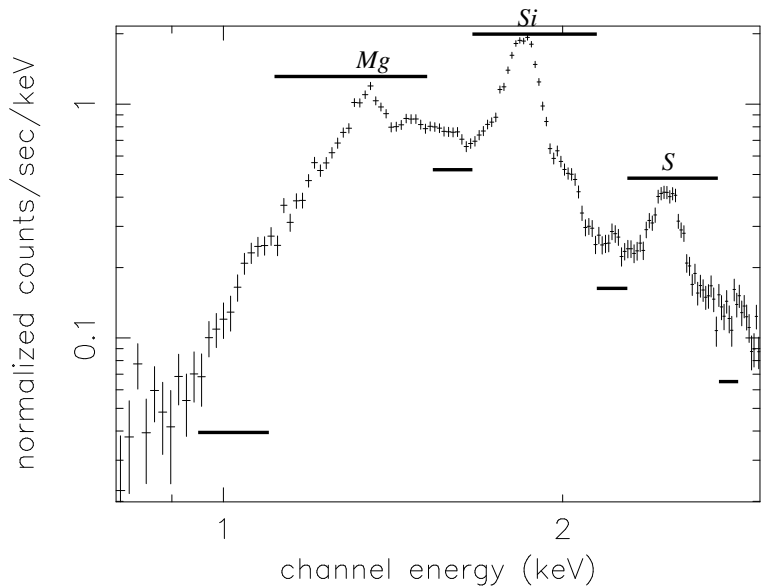,width=100mm,height=76.945mm,angle=0.0}
   \caption{ The {\sl Chandra} ACIS spectrum of the entire remnant.
 Horizontal bars above the spectrum, which are labelled by element, show the energy
   intervals used for each EW image; bars below the spectrum show the intervals
   used for the continua (see table 1).  }
   \end{center}
\end{figure}

\begin{table}[]
  \caption[]{ Energy Bands Used for Generating the
      Equivalent-Width Images
    }
  \label{Tab:publ-works}
  \begin{center}\begin{tabular}{clcl}
  \hline\noalign{\smallskip}
Elements  &   Line (eV)  &   Low (eV)\inst{a}   &   High (eV)\inst{a}                 \\
  \hline\noalign{\smallskip}
Mg(Fe~L)  &    1160-1560         &       950-1150       &   1570-1670 \\
Si        &    1680-2050         &       1570-1670      &   2100-2250 \\
S         &    2330-2730         &       2100-2250      &   2740-2840 \\  
  \noalign{\smallskip}\hline
\end{tabular}\end{center}
\inst{a} The low- and high-energy bands around the selected line energies
          used to estimate the underlying continua.


\end{table}

\begin{figure}
   \vspace{2mm}
   \begin{center}
   \hspace{3mm}\psfig{figure=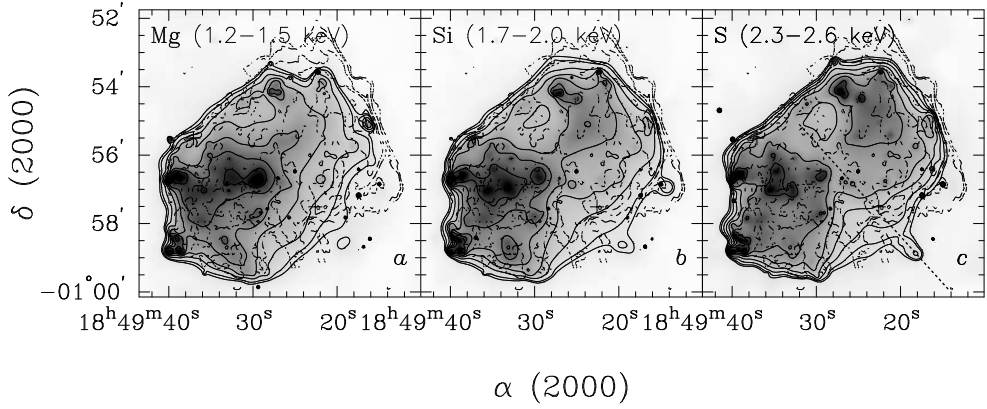,width=130mm,height=53.657mm,angle=0.0}
   \caption{
Adaptively smoothed narrow band 1.2-1.5, 1.7-2.0, and 2.3-2.6 keV
(including Mg He$\alpha$, Si He$\alpha$, and S He$\alpha$, respectively)
diffuse emission images (with S/N ratio of 3)
overlaid with the dashed contours of 1.5 GHz radio emission
(at 1.5, 3.19, 8.25, 16.69, 43.68, and $62.24 \times 10^{-3}$Jy/beam )
(Moffett \& Reynolds 1994).
The seven levels of solid contours are plotted with square-root
intensity scales between the maximum
and the 15\% maximum brightness. The two plus signs in each panel 
denote the OH maser points (Frail et al.\ 1996).
The dotted line in (c) roughly indicates the symmetrical axis of the SW protrusion.
}
      \label{Fig:3C 391 spectrum}
   \end{center}
\end{figure}

\begin{figure}
   \vspace{2mm}
   \begin{center}
   \hspace{3mm}\psfig{figure=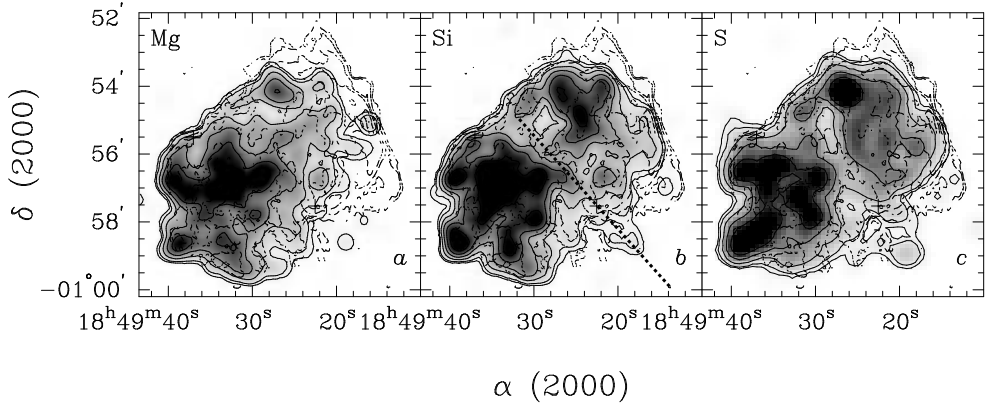,width=130mm,height=53.657mm,angle=0.0}
   \caption{
EW images of Mg, Si, and S lines. The Mg and Si (S)
images are extracted with $4"$ ($8"$) pixels and smoothed by a Gaussian
with $\sigma=12"$ ($16"$).
The seven levels of solid contours are plotted with square-root
intensity scales between the maximum and the 3\% maximum brightness.
These images are overlaid with the dashed contours (with the 
same levels as in Fig.\ 3) of 1.5 GHz radio 
emission (Reynolds \& Moffett 1993). The two plus signs in each panel 
denote the OH maser points (Frail et al.\ 1996).
The dotted line in (b) roughly indicates the symmetrical axis of the SW protrusion.
}
   \end{center}
\end{figure}
%
%

\section{Discussion}
\label{sect:discussion}

We have seen that both the narrow-band and EW images show similar
clumpy structure interior to the remnant and the finger-like
feature in Si and S protruding
radially outside the southwestern radio border. The gas clumps
have a normal abundance according to CSSW04 and thus are
essentially of interstellar origin.
The most intriguing phenomenon unveiled here is the
finger-like structure which is coincident with the
``spill-out'' portion seen in the broadband image (CSSW04).
Limited by the insufficient counts of the spectrum that is
extracted from the faint SW protruding region, we cannot make a
spectral fit to decide the metal abundances. However, the
similarity to the protruding feature of Si and other elements in
Cas~A may lend us a good clue to a possibility for the physical nature
of the finger-like feature in 3C~391.

The finger-like protruding feature on the SW seen in the Si and S EW
maps (Figs.\ 1c, 4b, and 4c) and in
the narrow-band maps (Figs.\ 3b, 3c) is, to some extent, similar
to the NE jet-like Si structure recently found in the famous SNR
Cas~A (Hwang et al.\ 2000). In Cas~A, the EW maps for Si
and other elements show a breakout protrusion on the NE, which is
explained as one of jets of ejecta extending outside of the blast
shock (Hwang et al.\ 2000, 2003, and 2004).
Hwang et al.\ (2004) further suggest that the
jets of  ejecta in Cas~A be yielded by an
asymmetric supernova explosion of a collapsar with normal
explosion energy ($2-4 \times 10^{51}$ ergs).
The asymmetric structures in Cas~A are not erased by the much less
massive hydrogen envelope at explosion compared to SN~1987A,
another core-collapse supernova
exploded asymmetrically (Wang et al.\ 2002 and references therein),
where the jets may have been erased by the hydrogen envelope (Chevalier \& Soker 1989).

The association of SNR 3C 391 with a dense molecular cloud (Wilner
et al.\ 1998; Reach \& Rho 1999) makes it very likely for this
remnant to result from an SN explosion of a massive core-collapse
progenitor star, although no stellar remnant has been found yet
(CSSW04).
The possibility for two point sources
(J184925.9$-$005628 and J184927.0$-$005640) located in the center
and a bright unresolved point-like source (J184922.3$-$005334), with
a power-law spectrum, near the NW boundary to be candidates of the
stellar remnant can not be ruled out (CSSW04). Thus we suggest
that the finger-like feature in Si and S lines in 3C~391 be
probably caused by the jet of ejecta of the massive progenitor.
Hence, like the case of Cas~A, there might only be low mass hydrogen
envelope at explosion of 3C~391, which has not erased the jet of ejecta.
Heger et al.\ (2003) and Chevalier (2005) suggest that stars with
a mass of 10-25$M_{\odot}$ will explode with most of there hydrogen
envelope present.
Therefore, if the jet-like structure in 3C~391 is true, the progenitor
star might have a mass not less than 25$M_{\odot}$.

Because of the irregular morphology of this remnant, the SN explosion
site has not been clear yet.
Reynolds \& Moffett (1993) suggest that the explosion site be located
within the northwestern half which is embedded by the dense cloud.
If the symmetric axis of the protruding feature could indicate the
trajectory of the ejection, then the explosion site may roughly be aligned
with it.
We plot such an axis in Fig.\ 3c and 4b and thus see that the explosion
site may be located slightly northwestward from the remnant's
geometrical center.

By comparison, the SW finger-like protrusion in SNR 3C 391
extends $3.5'$, about 8 pc
at a distance 8 kpc, from the explosion site or the geometrical center of the
remnant,
longer than the X-ray jets of ejecta in Cas A,
which measure exceeding 3.5$'$, about 3.5 pc at a distance 3.4 kpc
(Hwang et al.\ 2004).
The fast-moving ($\sim1.2\times10^{4}$ km s$^{-1}$) knots along the jets of Cas~A
reach as far as $4.5'$ ($\sim4.5$ pc) from the center (Fesen 2001).
If the protruding feature in 3C~391 indeed indicates a jet of ejecta,
this contrast of lengths could be reasonable in view of the
different ages and ambient gas densities of the two remnants.
3C~391 is estimated to be in an age of $\sim4\times10^{3}$ yr (CSSW04)
much larger than Cas~A's age $\sim 320$ yr (Fabian et al.\ 1980),
meanwhile one should note that the ambient gas density of 3C~391
($\sim30$ cm$^{-3}$) is much higher than that of Cas~A
($\sim3$ cm$^{-3}$) (Laming \& Hwang 2003).
One could naturally expect that the jets in Cas~A could extend longer than
4.5 pc at an older age with a considerable deceleration.

The nature of the finger-like feature, especially the
abundances of the metal species in this region, should be further
investigated with a deeper X-ray exposure of one of the space missions.

\section{Conclusion}
We have constructed the narrow-band and EW images of SNR 3C~391
for the emission lines of Mg, Si, and S using the {\sl Chandra}
ACIS observational data.
These images display a clumpy spatial distribution
of metal species Mg, Si, and S within an elongated periphery of the supernova
remnant, similar to the structure seen in the broadband images,
suggestive mainly of interstellar origin.
The most intriguing result is the finger-like protrusion revealed in Si and S
lines on the SW border.
A comparison is made between this protruding feature and the X-ray jets of ejecta
seen in lines of Si and other species in SNR Cas~A.
The similarity between these features may suggest a possibility
that the SW protruding finger-like feature in SNR 3C~391 be a trace of
the jet of ejecta of an asymmetric core-collapse supernova explosion of
a massive ($\gsim25M_{\odot}$) progenitor star.

\begin{acknowledgements}

We thank the anonymous referee for valuable advices. We also
thank Jasmina, Lazendic and Bing Jiang for the help in
the EW map production. 
This work is supported by NSFC Grants 10073003 and 10221001 and CMST Ascent Project Grant NKBRSF-G19990754.
\end{acknowledgements}

\label{lastpage}

\end{document}